# AGE DETERMINATIONS OF THE OPEN STAR CLUSTERS KING 14 AND NGC 146

**V. Kopchev, P. Nedialkov*, G. Petrov**

**Abstract**

We based on 2MASS J and $K_s$ photometry for the open star clusters King 14 and NGC 146, and using color magnitude diagrams with isochrones fit we have found an age of log(age) = 7.8 (63 ±8 Myr) for King 14 and log(age) = 7.5 (32 ±8 Myr) for NGC 146. Our age determination is bigger than given in Lynga [5] ( log(age) = 7.2 for King 14 and log(age) = 7.1 for NGC 146 ) and less than Dias [6] ( log(age) = 7.9 for King 14 and log(age) = 7.8 for NGC 146 ).

**Introduction.** Open star clusters are physically related groups of stars held together by mutual gravitational attraction. Therefore, they populate a limited region of space, typically much smaller than their distance from us, so that they are all roughly at the same distance. They are believed to originate from large cosmic gas and dust clouds in the Milky Way, and to continue to orbit the galaxy through the disk. Over 1600 open clusters are known in our Galaxy, and this is probably only a small percentage of the total population which is probably some factor higher, estimates of as many as about 100 000 Milky Way open clusters have been given. Most open clusters are young, generally less than a few hundred million years old. They are rich in the youngest and most heavy element-rich stars. Open clusters are very important objects in the study of stellar evolution, because the stars are all of very similar age and chemical composition, the effects of other more subtle variables on the properties of stars are much more easily studied than they are for isolated stars. The existence of double or binary clusters in the our neighbour galaxy Magellanic Clouds is fairly well established, whereas only one clusters pair h + $\chi$ Persei is known in our own Galaxy, Subramaniam at al. [1] suggested a catalogues of 18 probable binary open star clusters. For investigation these clusters since 1996 have started a project in collaboration between the Institute of Astronomy on Bulgarian Academy of Sciences and University of Bonn. This paper belongs to a series of papers where we will present our results. The aim of this study is to check and precise the ages for both probable binary open star clusters King 14 and NGC 146, using J and $K_s$ photometry from Two Micron All Sky Survey (hereafter 2MASS).

**Cluster data and age determinations.** Lindoff [2] is one of the first who derived a distance of 2440 pc and an age of 13 Myr for NGC 146, and 1960 pc and 16 Myr for King 14. Janes & Adler [3] quoted reddening and distances of 0.47 mag and 2800 pc for King 14, and 0.58 and 3300 pc for NGC 146. First CCD photometry for NGC 146 was presented by Phelps & Janes [4], who suggested a reddening of 0.70 mag, a distance of 4786 pc, and an age of less than 10 Myr. Lynga [5] quoted distance of 2600 pc, log(age) = 7.2, reddening E(B-V) = 0.55 for King 14 and 2900 pc, log(age) = 7.1, E(B-V) = 0.69 for NGC 146. Dias et al. [6] give the following fundamental parameters for King 14 and NGC 146 in the new catalogue of open clusters in the Galaxy Table 1.

Table 1
Basic data for King 14 and NGC 146

| | King 14 | NGC 146 |
|---|---|---|
| R.A. (2000) | 00 31 54 | 00 33 03 |
| Decl. (2000) | +63 10 00 | +63 18 06 |
| Ang. Diameter (arcmin) | 6.0 | 7.0 |
| Distance (pc) | 2593 | 3032 |
| E(B-V) (mag) | 0.414 | 0.480 |
| Log (age) | 7.924 | 7.822 |

We have used the photometry for the both clusters from 2MASS using VizieR tool available at http://visier.u-strasbg.fr. We have made circular extractions centered on the coordinates for each cluster ( King 14: $\alpha_{2000} = 00^h\ 31^m\ 54^s$ , $\delta_{2000} = +63^{\circ}\ 10'\ 00''$ and NGC 146: $\alpha_{2000} = 00^h\ 33^m\ 03^s$ , $\delta_{2000} = +63^{\circ}\ 18'\ 06''$ ). We have used for King 14 an extraction radius of 3.0 arcmin, and 3.5 arcmin for NGC 146. We have derived $E(J-K_s) = 0.22 \pm 0.02$ for King 14, and $E(J-K_s) = 0.25 \pm 0.02$ for NGC 146, using the interstellar extinction law: $A_J/A_V = 0.282$ and $A_K/A_V = 0.112$ from Rieke & Lebofsky [7], and data for the distance and the reddening from [6]. Colour-magnitude diagrams (CMDs) $M_J$ versus $(J-K_s)_0$ for clusters are given on Fig.1 for King 14 and Fig. 2 for NGC 146.

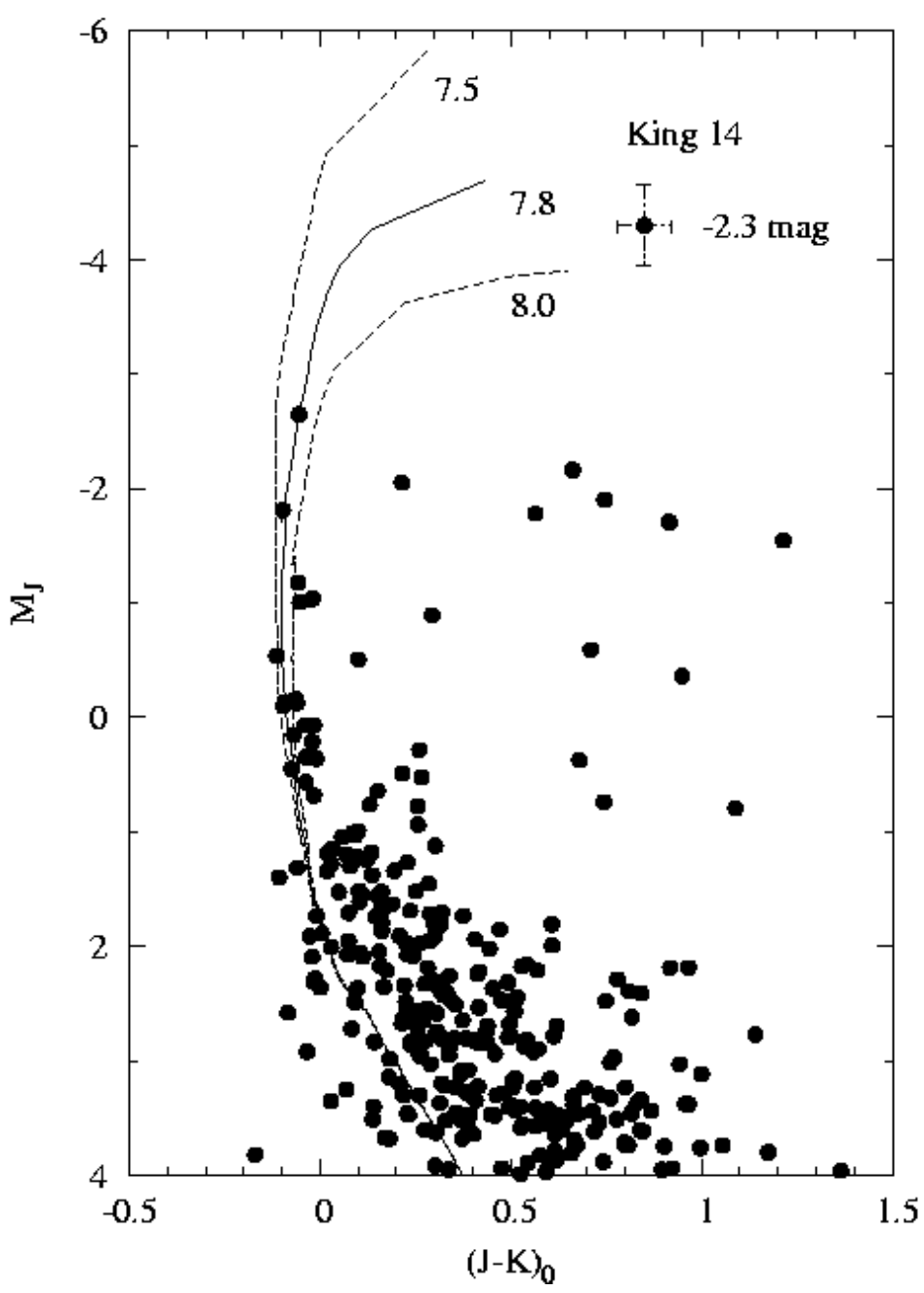

Fig. 1. Colour-magnitude diagram for King 14 with the best isochrone fit. We have adopted an age of log(age) = 7.8 (63 ±8 Myr).

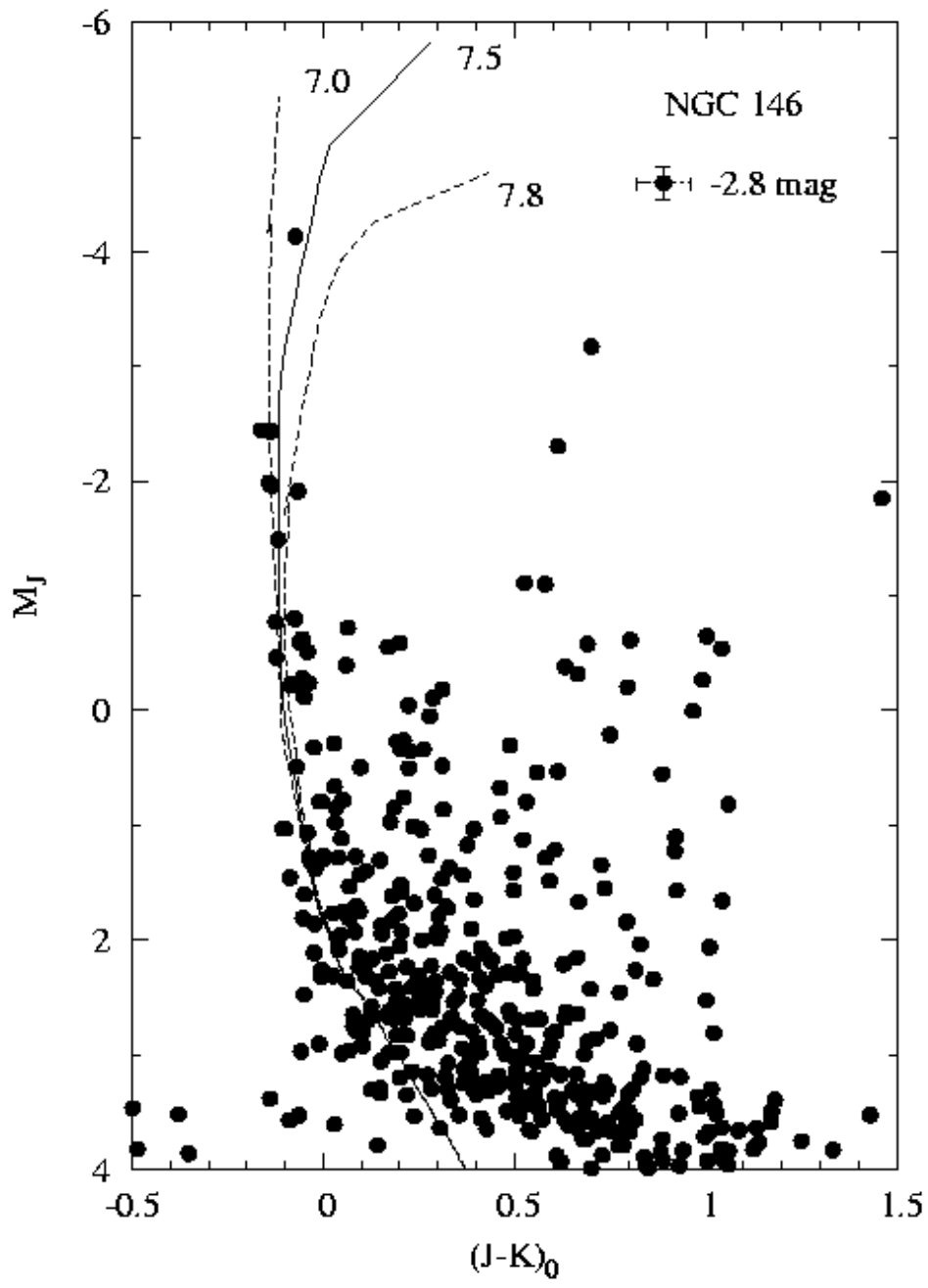

Fig. 2. Colour-magnitude diagram for NGC 146 with the best isochrone fit. We have adopted an age of log(age) = 7.5 (32 ±8 Myr).

We have determined the age of the clusters overplotting the best fitting isochrones on the CMDs. We have adopted an age of log(age) = 7.8 (63 ±8 Myr) for King 14, and log(age) = 7.5 (32 ±8 Myr) for NGC 146. The isochrones are based on the stellar models of the Geneva group Schaerer D.,at al. [8], whit Z = 0.008 which corresponds to metallicity [Fe/H] ≈ -0.3 dex.

**Conclusions.** We used 2MASS J and $K_S$ photometry for the two open star clusters King 14 and NGC 146, and fitting CMDs with isochrones based on the Geneva models, we have found log(age) = 7.8 (63 ±8 Myr) for King 14 and log(age) = 7.5 (32 ±8 Myr) for NGC 146. This is an age between the given age in Lynga [5] ( log(age) = 7.2 for King 14 and log(age) = 7.1 for NGC 146 ) and Dias [6] ( log(age) = 7.9 for King 14 and log(age) = 7.8 for NGC 146 ).

**Acknowledgements.** Our work is partially supported by the grant F-1302/2003 of the Bulgarian NSF.
This publication makes use of data products from the Two Micron All Sky Survey, which is a joint project of the University of Massachusetts and the Infrared Processing and Analysis Center/California Institute of Technology, funded by the National Aeronautics and Space Administration and the National Science Foundation.